# Faraday-cage screening reveals intrinsic aspects of the van der Waals attraction

Musen Li[a], Jeffrey R. Reimers[ab*] John F. Dobson[c], and Tim Gould[c*]

a: International Centre for Quantum and Molecular Structures and Department of Physics, Shanghai University, Shanghai 200444 China

b: School of Mathematical and Physical Sciences, University of Technology Sydney, NSW 2007 Australia

c: School of Natural Sciences and Queensland Micro- and Nanotechnology Centre, Griffith University, Nathan, QLD 4111, Australia

* email:  jeffrey.reimers@uts.edu.au, t.gould@griffith.edu.au.

**ABSTRACT**   General properties of the recently observed screening of the van der Waals (vdW) attraction between a silica substrate and silica tip by insertion of graphene are predicted using basic theory and first-principles calculations.  Results are then focused on possible practical applications, as well as an understanding of the nature of vdW attraction, considering recent discoveries showing it competing against covalent and ionic bonding.  The traditional view of the vdW attraction as arising from pairwise-additive London dispersion forces is considered using Grimme's "D3" method, comparing results to those from Tkatchenko's more general many-body dispersion (MBD) approach, all interpreted in terms of Dobson's general dispersion framework.  Encompassing the experimental results, MBD screening of the vdW force between two silica bilayers is shown to scale up to medium separations as 1.25 $d_e/d$, where $d$ is the bilayer separation and $d_e$ its equilibrium value, depicting *antiscreening* approaching and inside $d_e$.  Means of unifying this correlation effect with those included in modern density functionals are urgently required.

Materials in zero, one, two, and three dimensions of relevance to conceived future fabrication and electronics technologies are often held together by the van Waals dispersion force.[1] Methods of measuring and first-principles simulations of the free-energies of formation of such systems are becoming available.[2] Often the critical issues involve situations in which the forces holding systems together become non-additive, i.e., the interaction between two parts of a system is modulated by the presence of nearby matter, with dispersion and other aspects all contributing.[3,4] At short distances typical of chemical bonding, it is now being recognized that dispersion forces can sometimes compete with traditional chemical covalent and ionic bonding forces to control outcomes.[5-10] Related parallel work demonstrates how ionic forces can control typical scenarios associated with dispersion,[11] as well as scenarios in which general solvent effects including dispersion control structure.[12,13] Alternatively, at long distances, the Casimir dispersion effect becomes critical,[14,15] as well as other exotic phenomena associated with the wavelike nature of charge polarization in nanoscale objects.[4] While answers to each of the issues raised can be formed in isolation, a generally useful understanding the van der Waals force remains elusive. Indeed, how different computational methods perceive dispersion forces at long and short distances have been found to be uncorrelated, raising fundamental questions concerning the nature of the force.[7] What happens to long-range phenomena at van der Waals separations and then as chemical bond distances are reached will form a key part of future understanding.

To initiate discussion, we consider the extremely non-additive van der Waals interactions observed by Tsoi et al.[16] in systems involving a silica substrate, a silica AFM tip, and an intervening conducting graphene layer. The remarkable result from this work is that a large inter-object dispersion force was *switched off* by the insertion of graphene in between the objects. Here, the origin and basic properties of this effect are elucidated using first-principles computational methods applied to a model 2D system. Discussion is considered using the framework for understanding van der Waals phenomena developed recently by Dobson.[17]

Studies of 2D materials are currently very prevalent, with first-principle simulations providing powerful tools to facilitate understanding.[6,18-21] While dispersion interactions are critical for determining the structure and properties of such systems, the most commonly used method applied for first-principles materials simulations, density functional theory (DFT) using a conventional generalized-gradient approximation (GGA), improperly treats its contribution. As a result, a wide range of empirical correction schemes are commonly added to GGA calculations so as to produce a realistic description of the critical interactions.[20,21] The vdW dispersion interactions described by these schemes typically involve sums over inter-atomic interactions, each described by the London force,[17,22-24] with only small corrections. Related variants include replacing the atomic sums by electron-density integrals. In either case, these approaches are intrinsically pairwise additive, meaning that adding more atoms to the system just systematically increases the net dispersion interaction.

We select two widely applied types of methods to investigate the Tsoi et al. experiment[16] and its wider implications. First, we use pairwise-additive approaches based on "D3"-type schemes of Grimme that have achieved wide-ranging success,[2,25-28] particularly when applied to understand chemical van der Waals structures and energetics at equilibrium separations. These methods are also being found useful for understanding dispersion contributions down to chemical bond-length scales.[5,7,27,29] Second, we apply the more general many-body-dispersion (MBD) approaches of Tkatchenko and others that also have been shown to be widely successful,[4,20,30-37] especially when large objects or conducting objects are involved at both equilibrium and larger separations. Of general interest also is how different methods treat the van der Waals force during small excursions from equilibrium.[38] More significantly, the observation that the van der Waals force can be switched off severely challenges the traditional pairwise-additive point of view and hence provides an external reference frame for this discussion. Here, combining the results from the different computational methods reveal that the switching off of the van der Waals interaction is a *Faraday cage* effect. The model system used for these calculations is shown in Fig. 1 and contains two bilayer silicon sheets separated by vacuum from an interposed graphene layer.

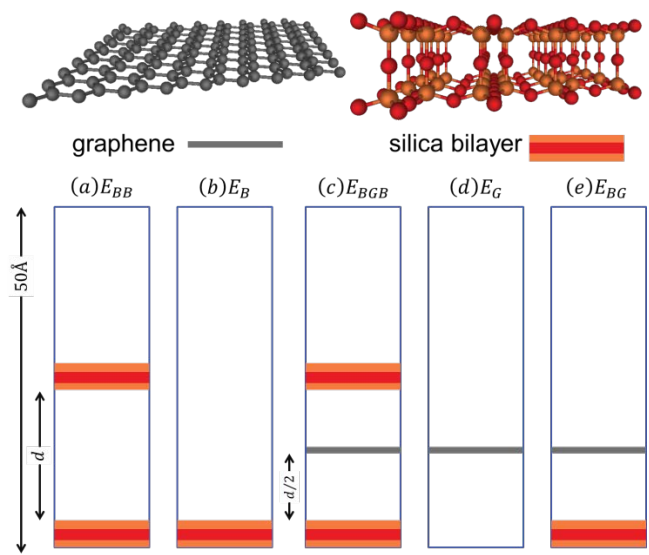

**Figure 1 | Geometries used for determining the screening of the vdW interaction** between two silica bilayers induced by inserting an intermediary graphene sheet.

Dobson's[17] framework for understanding the dispersion force conceptually links traditional ideas concerning pairwise additivity to "top-down" Lifshitz theory in which screening emerges as a natural consequence of the quantum fluctuations that facilitate the dispersion interaction.[39,40] In the traditional view, each inter-atomic interaction is specified by the London contribution[41] which scales as $r^{-6}$, where $r$ is the interatomic distance, with only small corrections.

The origin of the basic interaction is the polarization of one atom by spontaneous quantum dipole fluctuations occurring inside another.

At the simplest level, all atoms of the same type could be treated as having the same vdW attractions (the gas-phase values) independent of chemical environment. Dobson classified corrections to this approach into three types.[17] The first type, here called "Dobson-A", focuses on the quantal effect of the local chemical environment, insofar as it modifies the polarizability of each atom (e.g. via orbital compression) and hence modifies its vdW interaction with other atoms. Most modern computational methods treat such terms to high accuracy, sometimes through empirical parameterization and sometimes through explicit environmental modeling. In the second type, here called "Dobson-B", the fluctuating electric fields that mediate the vdW interaction between a pair of atoms are disrupted (screened or anti-screened) by the sympathetic induced fields caused by polarization of other atoms. This produces long-ranged "many-body" (many-atom) vdW effects.[17,23]. A third type of correction, called "Dobson-C", involves long-ranged fluctuating charge transfer occurring on a length scale larger than the size of an atom. This becomes particularly significant for *long-ranged* interactions between low-dimensional metallic conductors,[22,42,43] but, in undoped graphene at $T = 0$ K, it is less important.[18]

Hence, in the context of the Tsoi et al. experiment,[16] Dobson-B effects take on a central focus: (i) what are they? (ii) how reliably do modern computational methods treat them? (iii) can they account for the observed screening of the silica-silica dispersion attraction by graphene? (iv) how do they influence critical equilibrium properties of van der Waals heterostructures? (v) how do they influence the way that dispersion forces are perceived at shorter distances of the order of chemical bond lengths, (vi), how do they behave asymptotically at long range, and (vii) can they be manipulated to make new functional materials and devices?

**Results**

**Nature of the Dobson-B effect.** Figure 2 illustrates the significance of Type-B effects in the situation that we will model here, namely a monolayer graphene sheet midway between two well-separated silica sheets. The graphene sheet will turn out to reduce the dispersion interaction between the two $SiO_2$ sheets in a "Faraday-cage" effect. This effect arises within classical electrostatics and prevents an external electric field from penetrating through an infinite conducting sheet or a continuous metallic surface or "cage". Typical ramifications of this effect include the blocking of radio waves by a steel-framed bridge and the reduction of mobile-phone signals by typically a million-fold by placing a magnetic resonance imaging (MRI) machine inside a room with continuous metallic walls. Polarization of the atoms within the metal by the external field cooperate across the whole extent of the metal to prevent penetration of the electric and magnetic fields. The effect applies independent of the source of the electric field, e.g., permanent charges, electromagnetic radiation, or the spontaneous quantum dipole fluctuations involved in vdW attractions. Note, however, that the shielding of the fluctuating fields by an un-doped, cold graphene sheet should be less complete than the screening of static fields by a true metal sheet, but is still substantial, as will be demonstrated later.

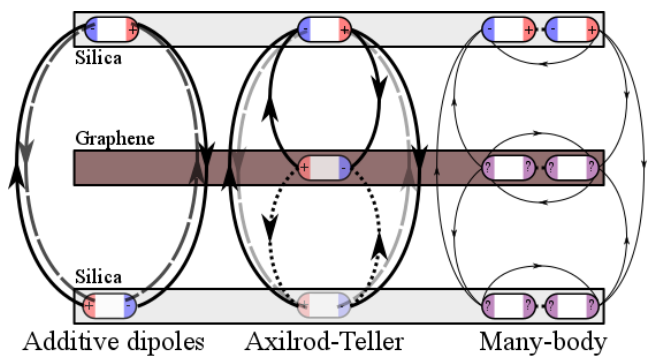

**Figure 2 | Many-body dispersion interactions.** Electric field lines (solid) produced by a short-lived spontaneous dipole on a top-layer atom induce polarization of other atoms: these atoms therefore produce reaction fields (dashed or dotted). Weaker fields and dipoles are here shown as fainter. See the text for a detailed analysis of the diagrams. A full many-body treatment was required for the "Faraday-cage" reported here.

*Pairwise dispersion theories (left diagram in Fig. 2)* give the direct $SiO_2$-$SiO_2$ dispersion energy via interaction of the original top dipole with the dashed reaction field generated by the induced bottom dipole.[44] This process misses simultaneous interactions involving both silica bilayers and graphene, and so cannot describe Faraday caging. The simplest way to include such interactions is through the *three-atom Axilrod-Teller interaction terms (middle diagram in Fig. 2)*. These arise because the solid field lines from the top dipole also induce dipoles in the graphene layer. A graphene dipole then produces the dotted field line, which tends to cancel the solid direct field line, at the position of the atom in the lower $SiO_2$ layer. This reduces the dipole induced on the bottom atom, and thereby reduces the dashed reaction field (fainter dashed lines). This in turn reduces the vdW interaction between atoms in the top and bottom layers. It is genuinely a 3-atom effect, additional to the pairwise interactions of all species. The middle atom has "screened" the pair interaction between top and bottom atoms, yielding a partial Faraday cage effect.

However, this Axilrod-Teller term is only the simplest level of the Dobson-B electrostatic screening phenomenon as unique contributions also appear when 4 atoms are simultaneously considered, etc.. *Many-body terms (right pane of Fig. 2)* lead to a more complete description of Faraday-cage screening. Here any atom can screen (or enhance, "antiscreen") the vdW interaction between any others, in accordance with the global geometry of the system. In this many-body case, for simplicity we do not distinguish initiating dipoles and fields from induced ones, showing all field lines as thin solid lines on the right-hand diagram of Fig. 2.

**Inclusion of Dobson-B effects in electronic structure calculations of molecules and materials.** Amongst ab initio computational approaches, the simplest that includes dispersion is second-order Møller-Plesset theory[45] (MP2). This includes dispersion up to Dobson-A only. Higher-order methods like coupled-cluster singles and doubles theory (CCSD)[46] and the random-phase approximation (RPA)[17] include elements of both Dobson-B and Dobson-C.

In DFT approaches, most treatments of dispersion, including double-hybrid functionals, are based intrinsically on the assumption of pairwise additivity of the dispersion energy,[17,21] thus excluding explicit Dobson-B effects. These include the "D3" family [except D3(ABC) which has some Axilrod-Teller terms],[25,26] the XDM method,[47] and many older approaches.[48-50] All practical computational methods involve many assumed equations and their parameters, with some methods (like D3) being fully empirical with many parameters, whilst others are semi-empirical, specifying equations that automatically generate required properties. Some methods may go beyond pure pairwise additivity by various means; D3(ABC) explicitly includes some 3-body contributions, whilst most other methods include an uncontrolled amount of Dobson-B contributions into the Dobson-A terms[21] by modifying the pairwise-additive energies based only on the immediate chemical environment.[51,52]

Going beyond this level of treatment, of the efficient, semi-empirical DFT+dispersion approaches, only the MBD method of Tkatchenko et.al[30,35] and its descendants[20,36] explicitly include *long-ranged* Dobson-B corrections to all orders. It does so in a way that depends on *global* geometries and so includes Faraday screening and other many-body effects. The RPA for DFT[44,53,54] includes all dispersive effects; efficient approaches of this type are emerging,[55-58] presenting a variety of possible options for future development. However, at the moment, only MBD and its descendants appear as feasible approaches for capturing screening effects in the complex chemical environment of the silica-graphene system. Demonstration that this approach can account for a large reduction of the vdW attraction between silica bilayers on insertion of graphene, whereas methods that do not fully include Dobson-B cannot, would present a result indicative that Faraday-cage screening is responsible for the analogous effect seen in the Tsoi et al. experiments.[16]

**Evidence for Faraday-cage screening by graphene.** The bilayer silica[16] used in our computational model constitutes a two-dimensional layered material that has similar properties to usual three-dimensional silica. However, it provides a test system small enough to allow calculations to very high precision (10 μeV), as is required for the evaluation of the dispersion force between bilayers over a large range in separations. A variety of specialized techniques are required to converge calculations to this precision, as detailed in the Supporting Information. To further enhance precision, the geometries of the silica bilayers and the graphene sheet were frozen at their individually optimized values. This is a poor approximation at and inside close contact, but is very adequate in the critical intermediate-range region. Similarly, we choose a commensurate lattice with lattice parameter 5.18 Å for silica and graphene. This preserves key qualitative properties of both materials, such as band structures, but ensures the cell size is reasonable; key results are insensitive to the value chosen.

To determine the dispersion interaction between two silica bilayers and the effect of intervening graphene on it, calculations are performed at 5 geometries, as described in Fig. 1. Figure 1(a) shows two silica bilayers separated by interfacial distance *d*; its energy is labelled $E_{BB}$. To get the unscreened dispersion energy $\Delta E_u$ at this distance, the energy of the corresponding isolated single bilayers, $E_B$, obtained from the structure shown in Fig. 1(b) is subtracted, yielding

$$\Delta E_u = E_{BB} - 2E_B \ . \tag{0}$$

To get the screened dispersion interaction $\Delta E_s$, we insert a graphene layer in the center between two silica bilayers, as shown in Fig. 1(c). The energy of this structure is labelled $E_{BGB}$, from which must be subtracted the energies of each component and the energies of each silica-graphene dispersion interaction. To do this, we evaluate the energy of an isolated graphene sheet $E_G$ (Fig. 1(d)) and the energy of the bilayer-graphene interaction at distances *d*/2, $E_{BG}$ (Fig. 1(e)), yielding

$$\Delta E_s = E_{BGB} - 2E_{BG} + E_G \tag{0}$$

Of particular interest is the ratio $\Delta E_s / \Delta E_u$ specifying the reduction of the inter-silica-bilayer dispersion force induced by the insertion of graphene. In addition, the total binding energy between the three layers is given by

$$\Delta E = E_{BGB} - E_G - 2E_B \ . \tag{0}$$

The D3 and MBD computational approaches both have multiple variants. For D3, we select the methods knows as "D3(BJ)" and "D3(BJ,ABC)", both involving Becke-Johnson damping, and where ABC indicates that some three-body Axilrod-Teller terms are included.[25,59] For MBD,[35,60] we employ the FI method,[37] which incorporates a superior polarizability model for oxides and a more physical treatment of many-body screening effects compared to the original method. Full details of the computational methods used are provided in Supporting Information.[61-65]

Figure 3(a) shows the total binding energy $\Delta E$ as a function of the interlayer spacing *d*, revealing that the predicted equilibrium separations are $d_e$ = 6.65 Å both for MBD-FI and D3(BJ,ABC) or else 6.50 Å for D3(BJ), all in the range reported by Wlodarczyk et al.[66] Next, Fig. 3(b) shows the unscreened vdW dispersion interaction $\Delta E_u$. All three methods considered give similar results for small *d*, with D3(BJ) predicting ~30 % larger values than the other methods at equilibrium and hence also ~30% larger binding $\Delta E$. While the magnitude of the additional binding strength predicted by D3(BJ) is significant, the overall similarity of the results reflects the observation that all of these methods can describe most systems in vdW contact.[2,20,21,67] Most significantly, the results from the very different approaches D3(BJ,ABC) and MBD-FI shown in Figs. 3(a) (for $\Delta E$) and 3(b) (for its small component $\Delta E_u$) are mostly in excellent agreement.

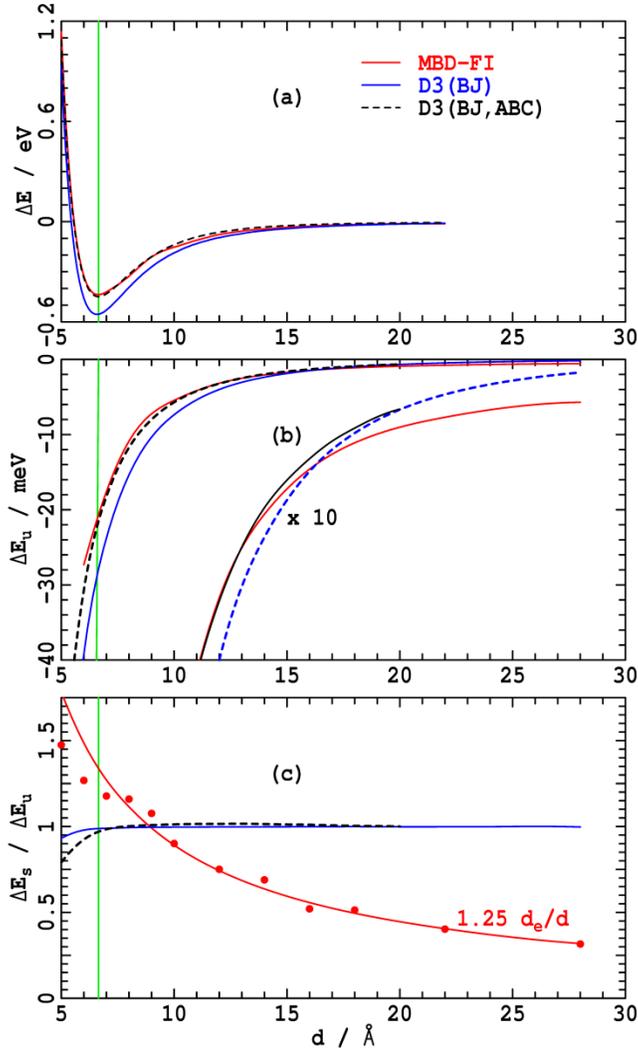

**Figure 3 | Screening at work.** (a) The total interaction energy $\Delta E$ at inter-layer distance $d$ (Fig. 1). (b) The contributions to this arising from the unscreened silica-silica van der Waals energies $\Delta E_u$ obtained without an intervening graphene. (c) The screening ratio $\Delta E_s / \Delta E_u$ obtained after insertion of graphene. The green lines indicates the equilibrium separation of $d_e = 6.65$ Å as calculated using MBD-FI and D3(BJ,ABC) ; D3(BJ) gives 6.60 Å instead.

However, how the various approaches treat the small bilayer-bilayer contribution to the total energy upon insertion of graphene, a quantity analogous to that measured in the Tsoi et al. experiments,[16] is highlighted in Fig. 3(c) where the ratio $\Delta E_s / \Delta E_u$ is shown. A purely pairwise-additive method would yield $\Delta E_s / \Delta E_u = 1$ at all geometries, and the results for both D3 variants are indeed very close to this value. Small deviations occur through explicit inclusion of 3-body corrections and the environment dependence of the parameters used in the method and, for larger distances, through rounding errors. In striking contrast, the MBD-FI ratio shown in Fig. 3(c) are fitted to yield

$$\frac{\Delta E_s}{\Delta E_u} = 1.25 \frac{d_e}{d}, \tag{0}$$

reducing towards zero with a $d^{-1}$ dependence. The screening is half the equilibrium value at twice the equilibrium bond length, but the value of 1.25 at equilibrium indicates that *antiscreening* (factor > 1) is occurring at this critical geometry. One feature relevant here is that perfect Faraday-cage screening only happens for continuous metallic conductors, and at short distances close to inter-atomic spacings, electric fields can penetrate through materials. Furthermore, the electron correlation effects manifested as the dispersion interaction must smoothly connect with the strong electron correlation effects manifested at short range within the GGA approximation, making the atomic nature of matter an unlikely cause of the antiscreening. In any case, a key feature of interest is that, at equilibrium distances, the net effect of Dobson-B terms becomes small (here a 25 % correction, with also a kink in the curve making the derivative small at this critical geometry), allowing sensibly parameterized pairwise-additive models to depict realistic many important features.[2,27,28,68]

However, the screening increases rapidly as the layers separate, reducing the unscreened van der Waals force to half at about twice the equilibrium separation. In this way, Dobson-B effects can become important even at or close to equilibrium separations, and approaches like MBD have been shown to have significant advantages in a range of applications.[4,20,30-37]

At moderately long range, extrapolation the MBD-FI results using Eqn. (3) predicts, at the separation of $d = 20$ nm as used in the Tsoi et al. experiments,[16] that the screening ratio $\Delta E_s / \Delta E_u$ should be 4.4 %, close to the observed value of 5 %. The calculations thus provide an explanation for the observation.

**Anticipated properties at long range.** The MBD calculations and the Tsoi et al. experiments[16] consider screening in the 0.5 – 3 nm range and at 20 nm, respectively. Anticipated properties asymptotically at long range are also worthy of consideration. As an example of Faraday cage screening, we note that a two-dimensional continuous metallic layer will completely reflect long-wavelength *static* ($\omega = 0$) electric fields, but even for a metal, *dynamic* (high-frequency) fields are not fully reflected. The dispersion interaction depends on the exchange of such *dynamic* (high $\omega$) electric fields. An undoped, zero-temperature graphene sheet is a good electrical conductor at zero frequency, but graphene is not a metal in the usual sense, having a zero electronic density of states at the Fermi energy. Correspondingly, it is known[69] that such a single undoped graphene layer, at $T = 0$ K as modelled herein, in contrast to a 2D metal, reflects most but not all of any long-wavelength *static* field, and this refection coefficient will be reduced at the finite frequencies of interest for the dispersion interaction. It is therefore not obvious how complete the Faraday caging effect will become for an undoped zero-temperature graphene sheet. The fitted $d^{-1}$ dependence (Eqn. 3) found at short to moderate distances may therefore in reality be replaced asymptotically by a constant minimum value

How the MBD method is expected to perform at very long range is also of interest. Rather than manifesting this anticipated theoretical limit, the MBD approach actually treats graphene as an insulator. Hence, asymptotically, MBD is expected to produce no screening at all (if it could be carried out numerically at asymptotic separations), and, like D3, would yield $\Delta E_s / \Delta E_u = 1$.

**Discussion**

Most chemical discussion of the van der Waals force considers it as an example of the effects of electron correlation operating at distances above 2-3 times chemical bond lengths. Table 1 summarizes understanding of its properties as well as more "chemical" properties associated with smaller length scales. The simplest approach for describing electronic structure, Hartree-Fock theory (HF), treats electrons as effective non-interacting particles, ignoring all electron correlation. It is qualitatively descriptive of covalent and ionic bonding but omits the subtle features often responsible for quantitative analysis of chemical reaction property differences and, accordingly, does not include any van der Waals effects.

**Table 1 | Overview of the effects of electron correlation.**

| Correlation type | example method | effect at chemical bond lengths | effect at near van der Waals separations | long-range Dobson-B effects like screening and antiscreening |
|---|---|---|---|---|
| None | HF | qualitatively descriptive of covalent and ionic bonds but poor quantitative accuracy | not included | not included |
| fully pairwise | MP2 | quantitatively useful | descriptive | not included |
| fully manybody | CCSD, RPA | quantitative | quantitative | quantitative |
| local manybody | GGA | quantitatively useful | improperly described | improperly described |
| local manybody + pairwise vdW | GGA+D3 | quantitativity improved by D3 | mostly quantitative | not included |
| local manybody + manybody vdW | GGA+MBD | quantitativity reduced by MBD | quantitative | adequate at short and intermediate range[a] |

a: except for low dimensional conductors for which Dobson-C becomes critical

Its ab initio improvement, MP2 theory, includes both short-range and long-range correlation in a pairwise fashion, greatly improving chemical reaction properties as well as describing van der Waals bonding, but Dobson-B effects like screening and antiscreening are not included. The excluded effects are well known to influence chemical bonding at short range, and here we see that they are also can be profound at long range. This shows that Dobson-B effects are influential over all length scales. At the next level of ab initio improvement to HF theory, methods like CCSD and RPA (approximately) include all effects, providing bonding descriptions to chemical accuracy as well as including screening and antiscreening.[17] While these methods do not directly provide insight in terms of intuitive descriptors like "chemical bonding" as distinct from "van der Waals bonding", they embody a *smooth link* between these concepts.

DFT approaches like GGA's with vdW corrections are in much more widespread use owing to their computational efficiency and their ready applicability to periodic systems, but the great challenge for the present concerns how to link together their disparate descriptions of "chemical bonding" and "van der Waals bonding". GGA's provide an exact description of electron correlation in a free-electron gas over all length scales and so in principle fully include dispersion. However, they do so in a way that yields very poor results in any atomic system, defining the core problem needing to be solved (Table 1). Of note is that they embody a fully manybody description of short-range electron correlation that typically provides a better description of chemical properties than does a purely pairwise method like MP2.

So the situation is that manybody effects are important at short range and, as demonstrated by the Faraday-cage screening effect, can also be critical at long range. Yet at intermediate distances typical of van der Waals equilibria, many-body effects appear to cancel, allowing the combination of GGA's with pairwise-additive schemes like D3 to become widely successful[2,27,28,68] (Table 1). Further, such schemes almost universally *improve* the quality of the description of the short-range chemical effects intrinsically provided by the GGA.[27] This effect becomes critical for the modern discussion of how van der Waals forces can remain in play down to the length scale of chemical bonds and possibly outcompete covalent and ionic bonding to control chemical structure.[6]

Table 1 indicates that replacing pairwise-additive methods like D3 with many-body ones like MBD allows all Dobson-B effects to be included at long range, often leading to improved descriptions at van der Waals contact.[4,20,30-37] One would hope that such improvements continued down to shorter length scales. Two examples illustrate key effects. In $ABP_2X_6$ monolayers, the internal chemical bonding has been shown to be dominated by dispersion forces acting within the monolayer, forces operative on the length scale of chemical bonds[8] that appear *uncorrelated* with the associated inter-layer forces.[7] In these systems, MBD gives relatively poor agreement with experiment whereas D3 appears moderately robust. By contrast, in interactions between benzene dimers, MBD significantly outperforms D3 at near-contact distances.[38]

These two cases, and the screening reported here, illustrate the competitive (and hard-to-define) role played by Dobson-A and -B terms at atomic scales, e.g., the effect of the horizontal field lines in the right diagram of Fig. 2 can by mimicked by local changes to the response function. These terms must meld smoothly into the description of electron correlation provided by the GGA. In D3, the Dobson-A terms do link smoothly to the GGA through the intuitive parameterization embodied in the method, a process that is less advanced in less empirical approaches like MBD, and even less advanced for the additionally included Dobson-B contributions in MBD.

Detailed information concerning how the Dobson-B terms need to be treated at short length scales is available in Fig. 3. The sharp rise of antiscreening at distances within the van der Waals separation predicted by MBD-FI is a significant feature that is not allowed by D3 and, by implication, may not be representative of real interactions.

**Conclusions**

Our results show that the dramatic switching-off of the van der Waals force demonstrated by the experiments of Tsoi et. al[16] is explained by Faraday-cage type screening effects involving the many-body response of all atoms in graphene to spontaneous quantum fluctuations in neighboring matter (Dobson-B effects). Modern computational methods can quantitatively reproduce these effects, provided that they fully include all responses to the fluctuations, as does MBD. However, subtleties, including correct asymptotic limits, needed to establish full quantitative agreement may require metallic (Dobson-C) models,[69] this becoming critical whenever fully metallic low-dimensional systems are involved.[17]

Methods like D3(BJ,ABC) that only partially include these responses are found to be inadequate for a general description of van der Waals screening in layered systems. These inadequacies are profound when screened silica bilayers are separated at distances of order twice the van der Waals separation and scale up with $d^{-1}$ at larger (but still intermediate) distances. However, the screening effect cancels as the equilibrium position is reached, allowing simplistic pairwise-additive methods to often yield good results for this important class of problems. We see here that the Tsoi et al. experiment[16] reveals fundamental physics critical to the understanding of systems at equilibrium as neglected screening (and other Dobson-B) effects appear as strong functions of separation around equilibrium structures. To better understand the dispersion force from asymptotic distances down to contact, there is great need for this experiment to be repeated so as to determine accurately the screening ratio $\Delta E_s / \Delta E_u$ as a function of distance $d$.

In terms of the conceptual understanding of the van der Waals force and its links to chemical forces, the results provide the first understanding as to how Dobson-B terms behave at short length scales below van der Waals equilibrium separations. They therefore provide insights into how one might improve van der Waals models at atomic bond-length scales via simultaneous improvements to Dobson-A and -B effects, tailoring atom-scaled van der Waals forces. An analytical understanding of how the RPA perceives Dobson-B effects at short range is urgently required.

In terms of nanotechnological applications, the consequences of Faraday-cage screening may become significant in future devices made from 2D materials. Already, multiple layered systems are being made, allowing screening to occur.[1] However, the screening effect acts to modify long-range interactions that are themselves small, so its influence may not be competitive with other effects in asserting control of self-assembly and device structure. This is in contrast to possible profound screening effects that conducting layers could have on functional properties housed inside layers that they separate, acting like a "ground plane" in electronics circuit boards.

To make the effects of van der Waals screening competitive, the objects being screened from each other need to themselves be large. The silica substrate and AFM tip used in the Tsoi et al. experiment[16] are examples of this effect. In this way, a conducting intermediary layer could dramatically reduce the force between two nanostructures of considerable thickness. If some external signal (e.g., transmitted current, applied electric or magnetic field, or an optical signal) could switch the conductance on or off, then the friction force between the nanoscale materials would also be modulated, blocking or facilitating slippage or separation.

The van der Waals force is not traditionally viewed as being controllable, as ionic forces, hydrogen bonding, and covalent bonding are. We now know that it can compete with these traditional chemical motifs to control structure and function.[6] Developing means of controlling and manipulating it therefore opens up new ways for designing functional materials and devices. The screening reported here shows that such manipulation is not only possible, but can be modelled using existing theories thereby enabling computer-led design of such materials.

## Acknowledgments

We thank Chinese NSF for support through Grant #11674212.

## Supporting Information

Details of the computational methods used.

**TOC graphic**

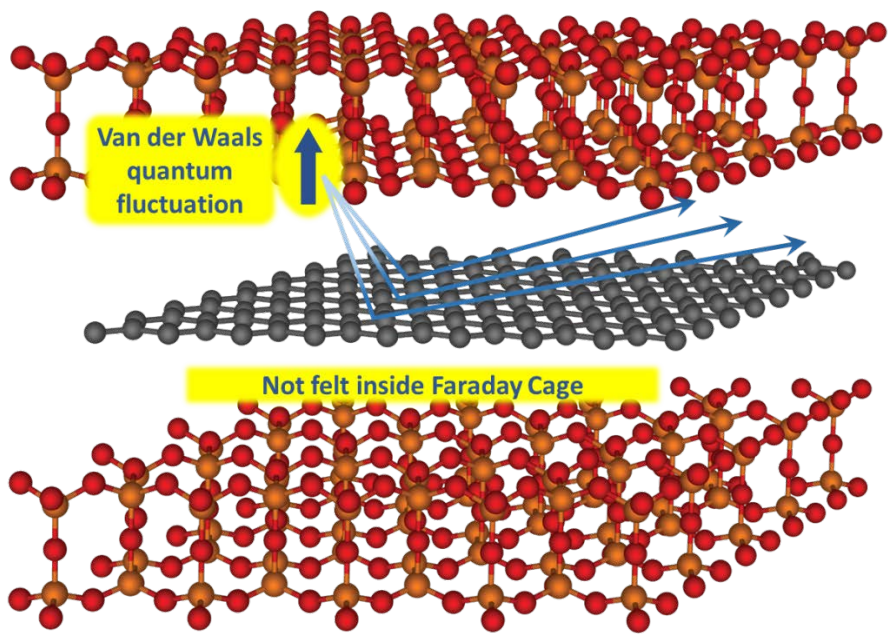